\pdfoutput=1
 \documentclass[aps,prl,reprint,twocolumn,showpacs,superscriptaddress,%
nobibnotes,
nofootinbib]{revtex4-1}  

\usepackage{color}
\usepackage{array}
\usepackage{graphicx}

\usepackage[english]{babel}
\usepackage[utf8]{inputenc}
\usepackage[tbtags]{amsmath}
\usepackage{amsfonts}
\usepackage{amssymb}
\usepackage{slashed}
\usepackage{graphicx}
\usepackage[T1]{fontenc}
\usepackage{ae}
\usepackage{color}
\usepackage[colorlinks=true,linkcolor=darkred,citecolor=darkred,urlcolor=darkred, pdfborder={0 0 0}]{hyperref}

\newcommand{\Eq}[1]{Eq.~(\ref{#1})}

\newcommand{\Fig}[1]{Fig.~\ref{fig:#1}}

\newcommand{\Tab}[1]{Table~\ref{tab:#1}}

\definecolor{gray}{gray}{0.1}
\definecolor{darkred}{rgb}{0.6471, 0.1098, 0.1882} 
\definecolor{darkblue}{rgb}{0.1098, 0.1882, 0.6471}

\def\beq{\begin{equation}}
\def\eeq{\end{equation}}

\def\beqn#1{\begin{equation}\label{#1}}
\def\eeqn{\end{equation}}

\def\beqa#1{\begin{eqnarray}\label{#1}}
\def\eeqa{\end{eqnarray}}

\def\gsim{\raise0.3ex\hbox{$\;>$\kern-0.75em\raise-1.1ex\hbox{$\sim\;$}}}
\def\lsim{\raise0.3ex\hbox{$\;<$\kern-0.75em\raise-1.1ex\hbox{$\sim\;$}}}

\def\calG{\mathcal{G}}

\def\calP{\mathcal{P}}

\definecolor{nicered}{rgb}{0.6,0.1,0.1}
\definecolor{applegreen}{rgb}{0.55, 0.71, 0.0}

\begin{document}

 \baselineskip 16pt

\title{Resonant production of dark photons in positron beam dump experiments}

 \def\LNF{\normalsize \it INFN, Laboratori Nazionali di Frascati, C.P.~13, I-00044 Frascati, Italy}
\def\UdeA{\normalsize\it  Universidad de Antioquia, Instituto de
                          F\'{i}sica, Calle 70 No. 52-21, Medell\'{i}n, Colombia}
\def\RMTRE{\normalsize\it Dipartimento di Matematica e Fisica,
                        Universit\`a di Roma Tre,  I-00146 Rome, Italy}
\def\INFNTRE{INFN, Sezione di Roma Tre, 
                 I-00146 Rome, Italy}
\def\RM{\normalsize\it Dipartimento di Fisica, Universit\`a di Roma La Sapienza 
                        and INFN, Sezione di Roma, 
                       I-00185 Rome, Italy}
 \author{Enrico Nardi}
 \email[Corresponding author, email: ]{enrico.nardi@lnf.infn.it}
 \affiliation{\LNF}
 \author{Cristian D. R. Carvajal}
\affiliation{\UdeA}
 \author{Anish Ghoshal}
\affiliation{\LNF}
\affiliation{\RMTRE}

 \author{Davide Meloni}
 \affiliation{\RMTRE}
 \affiliation{\INFNTRE}


%
\author{Mauro Raggi}
 \affiliation{\RM}

\date{\today}

\begin{abstract}
\noindent
Positrons beam dump experiments have unique features to search for
very narrow resonances coupled superweakly to $e^+ e^-$ pairs.  Due to
the continue loss of energy from soft photon bremsstrahlung, in the
first few radiation lengths of the dump a positron
beam 
can continuously scan for resonant production of new resonances via
$e^+$ annihilation off an atomic $e^-$ in the target.
In the case of a dark photon $A'$ kinetically mixed with the photon,
this production mode is of first order in the electromagnetic coupling
$\alpha$, and thus parametrically enhanced with respect to the
$O(\alpha^2)$ $e^+e^- \to \gamma A'$ production mode and to the
$O(\alpha^3)$ $A'$ bremsstrahlung in $e^--$nucleon scattering so far 
considered.  If the lifetime is sufficiently
long to allow the $A'$ to exit the dump, $A' \to e^+e^-$ decays could
be easily detected and distinguished from backgrounds.  We explore the
foreseeable sensitivity of the Frascati PADME experiment in searching
with this technique for the $17\,$MeV dark photon invoked to explain
the $^8$Be anomaly in nuclear transitions.
  \end{abstract}




\maketitle

\section{Introduction} 
\label{sec:intro} 

Some unquestionable experimental facts, like dark matter (DM),
neutrino masses, and the baryon asymmetry of the Universe, cannot be
accounted for within the standard model (SM) of particle physics. 
Physics beyond the SM (BSM) is thus required, which might
correspond to a whole new sector containing new particles as
well as new interactions.  If such a sector exists, there are two
possible reasons why it has not been discovered yet: (i) the mass
scale of the new particles, including the mediators of the new forces,
is well above the energy scale reached so far in laboratory
experiments; (ii) the mass scale is within experimental reach, but the
couplings between the new particles and the SM are so feeble that the
whole new sector has so far remained hidden.

The first possibility keeps being actively investigated mainly in
collider experiments, with the current high energy frontier set by the
LHC experiments.  However, the so far unsuccessful search for new
heavy states has triggered in recent years an increasing interest in
the second possibility, with many proposals and many new ideas to hunt
for new physics at the intensity frontier (see~\cite{Alexander:2016aln,Battaglieri:2017aum}
for  recent reviews).  In particular, the so called dark-photon (DP) or
$A'$-boson, that is a massive gauge boson arising from a new $U(1)'$
symmetry, can be considered as a natural candidate for a superweakly
coupled new state, since its dominant interaction with the SM
sector might arise solely from a mixed kinetic term
$(\epsilon/2) F'_{\mu\nu} F^{\mu\nu}$ coupling the $U(1)'$ and QED
field strength tensors, with values of $\epsilon$ 
naturally falling in a range well below $10^{-2}$.

From the phenomenological point of view, light weakly coupled new
particles have been invoked to account for discrepancies between SM
predictions and experimental results, as for example the measured
value of the muon anomalous magnetic moment~\cite{Blum:2013xva}, the
value of the proton charge radius as measured in muonic
atoms~\cite{Pohl:2010zza,Carlson:2015jba,Krauth:2017ijq,Pohl1:2016xoo},
or the anomaly observed in excited $^8$Be nuclear decays by the Atomki
collaboration~\cite{Krasznahorkay:2015iga,Krasznahorkay:2017gwn,Krasznahorkay:2017qfd}.
This last anomaly is particularly relevant for the present paper since
the new experimental technique that we are going to describe appears
remarkably well suited to test, at least in some region of the
parameter space, the particle physics explanation 
involving a new gauge boson with mass $m_{A'} \sim 17\,$MeV kinetically
mixed with the photon~\cite{Feng:2016ysn}.

The anomaly consists in the observation of a bump in the opening angle
and invariant mass distributions of electron-positron pairs produced
in the decays of an excited $^8$Be
nucleus~\cite{Krasznahorkay:2015iga}, which seems unaccountable by
known physics. The anomaly has a high statistical significance of
$6.8 \sigma$ which excludes the possibility that it arises as a
statistical fluctuation.  The shape of the excess is remarkably
consistent with that expected if a new particle with mass
$m_{A'} =17.0\pm 0.2{\rm (stat)}\pm 
0.5{\rm (sys)}\,$MeV~\cite{Krasznahorkay:2017qfd} is being produced in
these decays.
The strength of the $A'$ coupling to $e^+ e^-$ pairs, parametrized as
$\epsilon = \sqrt{\alpha'/\alpha}$ with $\alpha'$ the $U(1)'$ fine
structure constant, is constrained by different experimental
considerations.  In the Atomki setup, $A'\to e^+e^-$ decays must occur
in the few cm distance between the target, where the $^8$Be excited
state is formed, and the detectors. This implies a lower limit
$\epsilon/\sqrt{{\rm Br}(A' \to e^+e^-)} \gsim 1.3 \times 10^{-5}$ (we
will always quote limits on $\epsilon$ leaving understood that they
apply to its absolute value).  In the following we will assume for
simplicity ${\rm Br}(A' \to e^+e^-)=1$, if the $A'$ decay with a
non-negligible rate into invisible “dark” particles $\chi$, with
$m_\chi < m_{A'}/2$, the quoted limits need to be accordingly
rescaled. However, in case the invisible decay channel becomes largely
dominant, other limits different from the ones discussed in this paper
apply. We refer to Ref.~\cite{Izaguirre:2014bca} for details.

Lower limits on $\epsilon$ much stronger than what implied by the
Atomki experimental setup are obtained from electron beam dump
experiments. Old data from KEK~\cite{Konaka:1986cb} and
ORSAY~\cite{Davier:1989wz} have been reanalyzed in
Ref.~\cite{Andreas:2012mt} yielding, in the interesting mass range
$m_{A'} \sim 17\,$MeV, $\epsilon \gsim 7 \times 10^{-5}$.  A stronger
limit, $\epsilon \gsim 2 \times 10^{-4}$ was obtained in
\cite{Bjorken:2009mm} from a reanalysis the E141 experiment at SLAC
\cite{Riordan:1987aw}. However, for a $m_{A'}\sim 17\,$MeV the
excluded region is very close to the kinematic limit of the
sensitivity (see \Fig{limits}) and it has been recently pointed out,
by direct comparison with exact calculations~\cite{Liu:2017htz}, that
the Weizs\"aker-Williams (WW)
approximation~\cite{vonWeizsacker:1934nji,Williams:1935dka,Kim:1973he}
adopted to derive the limits become inaccurate in this kinematic
region, tending to overestimate the reach in
mass~\cite{Liu:2017htz,Gninenko:2017yus,Banerjee:2017hhz}. More in
detail, for primary energies in the the range $10-20\,$GeV, as was the
case for the E141 beam~\cite{Riordan:1987aw}, and for
$m_{A'} \sim 20\,$MeV, the WW approximation yields an $A'$ production
cross section about $50\%$ larger than the exact calculation (see
Fig. 2 in Ref.~\cite{Banerjee:2017hhz}) and it also overestimates the
$A'$ emission spectrum at large energies (see Fig. 4 in the same
reference), in which case the number of expected positrons falling
within the 1.1mrad angular acceptance of the experiment would be
overestimated both because of the larger boost, and also because of
the larger lifetime dilation that would cause the $A'$ to decay closer
to the detector.  Besides this, let us note that an $A'$ slightly
heavier than the benchmark value of 17\,MeV would in any case evade
the E141 limit.  It is then questionable if, for
$m_{A'}\gsim 17\,$MeV, the E141 constraints on the $A'$ couplings can
be considered as firmly established.  Conservatively, we will assume
that the corresponding region is still viable.

Upper bounds on $\epsilon$ in the relevant $A'$ mass range also exist,
see \Fig{limits}.
The KLOE-2 experiment has searched for $e^+ e^-\to \gamma A'$ followed
by $A' \to e^+e^-$ setting the limit
$\epsilon < 2 \times 10^{-3} $~\cite{Anastasi:2015qla}, while
constrains from the anomalous magnetic moment of the electron~\cite{Pospelov:2008zw}
yield $\epsilon < 1.4 \times 10^{-3}$~\cite{Endo:2012hp,Davoudiasl:2014kua}.  
A comparable limit stems from BaBar searches for $A'\to e^+e^-$
decays,  but it only applies for $m_{A'}> 20\,$MeV~\cite{Lees:2014xha}.
In summary, we will take the interval
\begin{equation}
  \label{eq:interval}
   7 \times 10^{-5} \leq \epsilon \leq 1.4 \times 10^{-3} \,.
\end{equation}
as the window allowed for a 17 MeV $A'$ decaying dominantly into
$e^+e^-$. This corresponds to a DP width $2.0\times 10^{-4}
\leq \Gamma_{A'}/{\rm eV} \leq 8.1\times 10^{-2} $.
%
\begin{figure}[t!]
\begin{center}
\includegraphics[width=0.80\linewidth,height=10cm]{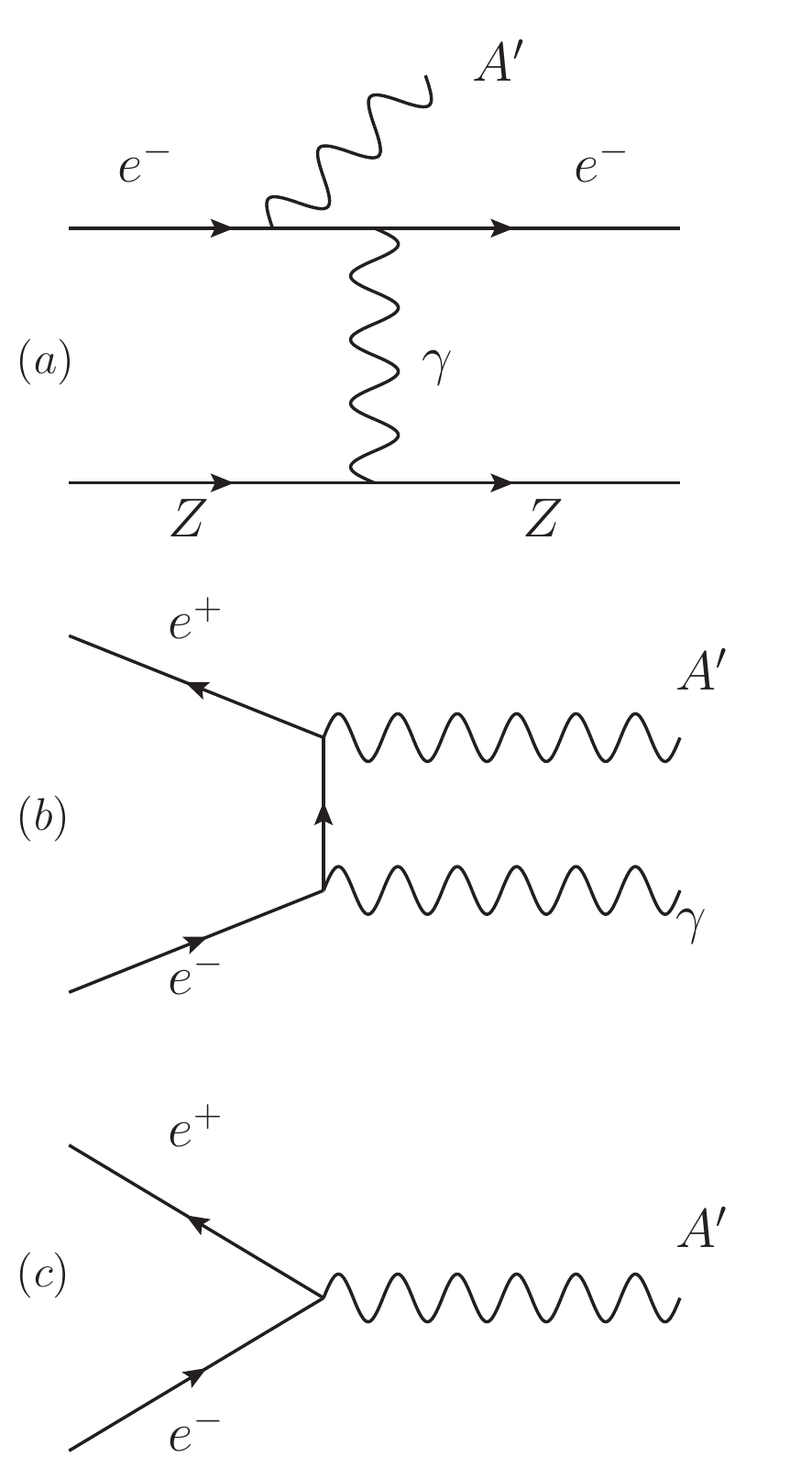}
\caption{$A'$ production modes in fixed target electron/positron
 beam experiments: 
  $(a)$~$A'$-strahlung in $e^-$-nucleon scattering;
  $(b)$~$A'$-strahlung in $e^+e^-$ annihilation; 
  $(c)$~resonant $A'$ production in $e^+e^-$ annihilation.}
\label{fig:fig-mech}
 \end{center}
\end{figure}
%
%

\section{The PADME experiment at LNF} 
Collider searches for dark photons have been carried out in electron
beam dump experiments (see \cite{Andreas:2012mt} for a review)
assuming $A'$-strahlung as the leading production mechanism in
electron-nucleon scattering.  Parametrically, this process is of order
$\alpha^3$, see \Fig{fig-mech}(a).  As regards $A'$ searches with
positron beams, there are only few facilities which, in the next
future, will be able to provide beams suitable for fixed target
experiments, and correspondingly only a few experimental proposals
have been put
forth~\cite{Rachek:2017gdc,Alexander:2017rfd,Raggi:2015gza}.  The
production mechanism considered so far is analogous to the usual QED
process of positron annihilation off an atomic target electron with
two final state photons, where one photon is replaced by one $A'$ see
\Fig{fig-mech}(b), corresponding to a process of $O(\alpha^2)$.  This
is the specific production process envisaged for the Frascati PADME
experiment~\cite{Raggi:2015gza} that we will now describe briefly.
\begin{table}[t!]
\setlength{\tabcolsep}{5pt} \global\long\def\arraystretch{1.2}
\begin{centering}
\begin{tabular}{|c|c||c|c|}
\hline 
   & pot/yr & $E_{\rm min}$ (MeV)   & $E_{\rm max}$ (MeV)  \\
\hline
\hline
 $e^+$ & $10^{18}$ & $ 250$ & $550$ \\
\hline
 $e^-$ & $10^{18} $& $ 250$ & $800$ \\
\hline
\end{tabular}
\par
\end{centering}
\caption{\label{tab:BTF} Beam parameters for the Frascati BTF.}
\end{table}

The PADME experiment \cite{Raggi:2014zpa,Raggi:2015gza} at the
DA$\Phi$NE LINAC Beam Test Facility (BTF)~\cite{Ghigo:2003gy} of the
INFN Laboratori Nazionali di Frascati (LNF) has been designed to
search for DP by using a positron beam~\cite{Valente:2017cko}
impinging on a thin target of low atomic number. The $A'$ can be
detected in the invisible channel by searching for a narrow bump in
the spectrum of the missing mass measured in single photon final
states, originated via $e^+e^-\to \gamma A'$.  The experiment will use
a 550 MeV positron beam impinging on a 100$\mu$m thick active target
made of polycrystalline diamond ($Z=6$). To keep under control the
counting rates the beam intensity will be kept at $\sim 10^{13}$
positrons on target per year (pot/yr), that is well below the maximum
available intensity (cfr.~\Tab{BTF}).  The low $Z$ and very thin
target are intended to minimize the probability of photon interaction
inside the target since, in order to reconstruct accurately the
missing mass, the measurement requires a precise determination of the
four-momentum of the $\gamma$ produced in the annihilation.  The
recoil photons will be detected by a quasi cylindrical calorimeter
made of inorganic crystals
located 3.3~m downstream the target, while the non-interacted
positrons, which constitute the vast majority of the incoming
particles, are deflected outside the acceptance of the calorimeter by
a 1\,m long dipole magnet.  Three different sets of plastic
scintillator bars will serve to detect electrons and positrons.
Profiting by the presence of a strong magnetic field, these detectors,
intended to provide an efficient veto for the positron bremsstrahlung
background, can also be used to measure the charged particles
momentum. The PADME detector is thus able to detect photons and
charged particles and it will be sensitive to invisible
($A'\to \chi\bar{\chi}$) as well as to visible ($A'\to e^+e^-$) DP
decays. PADME will start taking data already during May 2018.

\section{$A'$ production via\ resonant $e^+e^-$ annihilation}
\label{sec:production}
In this Letter we point out that for $A'$ masses $\gsim 1\,$MeV, the
process of resonant $e^+e^-$ annihilation into on-shell $A'$ depicted
in \Fig{fig-mech}(c), represents another production mechanism which,
being of $O(\alpha)$, is parametrically enhanced with respect to the
previous two production channels. Besides this, $A'$ production via
resonant $e^+e^-$ annihilation has several other advantages that we
will illustrate below, which altogether suggest that it might be
particularly convenient to operate the PADME (as well as other)
positron beam fixed target experiment in a dedicated mode in order to
search for $A'$ via resonant production.  Besides experiments with
positron beams, resonant $e^+e^-\to A' $ annihilation must also be
accounted for in a correct analysis of electron beam dump experiments
since, as is remarked in \cite{Marsicano:2018krp}, positrons are
abundantly produced in the electromagnetic (EM) showers inside the
dump.  This feature was recently exploited in~\cite{Marsicano:2018krp}
in reanalysing old results from the SLAC E137
experiment~\cite{Bjorken:1988as} by including $A'$ production via
resonant annihilation (and, but less importantly, also $A'$-strahlung
in annihilation). As a result, it was found that due to the
contribution of resonant $A'$ production, the E137 data exclude a
parameter space region larger than it was previously
though~\cite{Bjorken:2009mm,Andreas:2012mt}.  The extended excluded
region corresponds to the area in light grey color towards the bottom
of the plot in~\Fig{limits}.  Hence, in analysing electron beam dump
data, $A'$ production from annihilation of secondary positrons via the
diagrams in \Fig{fig-mech}(b) and (c) should be also accounted for.

In this section we consider the sensitivity of the PADME experiment to
the production process $e^+ e^- \to A' \to e^+e^-$.  In order to
exploit the resonant production mechanism, however, an experimental
setup slightly different from the one originally conceived is more
convenient.  The thin diamond target should be replaced by a tungsten
target of several cms of length, and this for two main reasons. The first one
is that of absorbing most of the incoming positron beam and of the
related EM showers, and in any case to degrade sufficiently the energy
of the residual emerging particles, so that the charged particles 
background can be easily deflected and disposed of. The $A'$
produced in $e^+e^-$ annihilation, if sufficiently long lived, will
escape the dump without interacting, and will decay inside the
downstream vacuum vessel, producing an $e^+e^-$ pair of well defined
energy.  The thick tungsten target thus allows to take advantage of
the full beam intensity of $10^{18}\,$pot/yr, with a gain of five
orders of magnitude with respect to the thin target running mode, see
\Tab{BTF}~\footnote{The maximum number of $e^\pm$ deliverable in one
  year given in the table (the one we will use) is LNF site
  authorization limited by the efficiency of the existing radiation
  shielding. However, technically the BTF could deliver up to
  $10^{20}\,$ electrons or positrons on target per year.}.  The second
reason for using a thick target is that of providing an almost
continuous energy loss for the incoming positrons
propagating through the dump, so that they can efficiently `scan' in
energy for locating very narrow resonances.

The energy distribution of positrons inside the BTF beam, tunable to a
nominal energy $E_b$ within the range $250\leq E_b/{\rm MeV} \leq 550$, can
be described by a Gaussian $\calG(E) = \calG(E;E_b,\sigma_b)$ where
$\sigma_b/E_b \sim 1\% $ is the energy spread.  The probability that a
positron with initial energy $E$ will have an energy $E_e$ after
traversing $t=\rho\cdot z/X_0$ radiation lengths (with $\rho$ the density
of the material in g/cm$^{-3}$ and $X_0 = 6.76\,$g/cm$^{-2}$ the unit
radiation length in tungsten), is given by
\cite{Bethe:1934za,Tsai:1966js}
  \begin{equation}
    \label{eq:radlength}
    I(E,E_e,t) = \frac{\theta(E-E_e)}{E\, \Gamma(bt)}
    \left[\log\frac{E}{E_e}\right]^{bt-1}, 
  \end{equation}
  where $b=4/3$ and $\Gamma$ is the gamma function. \Eq{eq:radlength}
  neglects secondary positrons from EM showers, as well as the loss of
  primary positrons from $e^+e^-\to \gamma\gamma$ annihilation, but is
  still sufficiently accurate for our purposes.  The $e^+$ energy
  distribution after $t$ radiation length is given by:
  \begin{equation}
    \label{eq:F}
    \mathcal{T}(E_e,t)= \int_{0}^\infty
\calG(E)\;    I(E,E_e,t)\; dE \,.
  \end{equation}
  Integrating $\mathcal{T}(E_e,t)$ in $t$ one would obtain the
  track-length distribution for primary positrons.  However, for an
  accurate determination of the detectable number of $A'$, the
  coordinate $z=t X_0/\rho$ of the production point is important,
  especially for the larger $\epsilon$, and hence shorter decay lengths. Thus,
  the integration in $t$ should be performed only when accounting for 
  the probability of $A'$ decaying outside the dump.  We fix the origin
  of the longitudinal coordinate at the beginning of the dump, $z_D$
  is the end point of the dump, and $z_{\rm det} $ is the distance
  between the origin and the detector. The $A'$ decay length
  $\ell_\epsilon = c\,\gamma \tau_{A'}$, with
  $\gamma = \frac{m_{A'}}{2 m_e}$ the time dilation factor, depends
  quadratically on $\epsilon $ through the lifetime
  $\tau_{A'}=1/\Gamma_{A'}$ (but it does not depend on $m_{A'}$, see
  below).  For the range of $\epsilon$ given in \Eq{eq:interval},
  $16 \gsim \ell_\epsilon/{\rm mm} \gsim 0.04$.  The number of
  detectable DP events then is:
\begin{equation}
  \label{eq:observable} 
N_{A'}
=\frac{N_{e^+} N_0 X_0 Z}{A} 
e^{-\frac{z_D}{\ell_\epsilon}}\! 
 \int_0^{T} \!\!\! dt\,      
e^{\frac{X_0}{\rho \ell_\epsilon} t} \!
 \int_0^{\infty} \!\!\!\!dE_e \, \mathcal{T}(E_e,t)\,     
\sigma_{\rm res}(E_e)\,,  
\end{equation}
with $N_{e^+}$ the number of incident positrons, $N_0$ the Avogadro
number, $A=184$ the atomic mass of tungsten, $Z=74$ is the atomic
number and
$\sigma_{\rm res}(E_e)$ the differential resonant cross section.
\Eq{eq:observable} takes into account the fact that the probability to
detect an $A'$ produced at $z$ is given by the integral of
${d \calP}/{dz} = ({1}/{\ell_\epsilon}) e^{-z/\ell_\epsilon}$ between
$z_D-z$ and $z_{\rm det} \to \infty$, where the limit is justified
since $z_D\sim O(1\,{\rm m})$).  Moreover, if the initial beam energy
happens to be not much above the resonance, after just a fraction of a
radiation length ($\rho X_0=3.5\,$mm for tungsten) the energy of most
positrons will have already degraded below the threshold for resonant
production, so that setting $T=1$ for the upper limit of the
integration is also a good approximation.  In \Eq{eq:observable} the
first exponential accounts for the fact that the larger is the length
of the dump, the smallest is the number of $A'$ that can be
detected. For $z_D\sim 10\,$cm we can expect that virtually all the
background from the EM showers will be absorbed in the dump. However,
only a few $A'$ will decay outside. To increase the statistics we can
reduce $z_D$, but then keeping the background under control can become
an issue.  In the lack of a dedicated simulation of the
detector/background for the resonant annihilation process, we will
estimate the sensitivity to the $A'$ couplings that could be achieved
with $z_D = 10\,$cm, $z_D=5\,$cm, and $z_D=2\,$cm (in the last two
cases a reduction of the beam intensity to keep under control
background contamination might be required).
As regards  $\sigma_{\rm res}$, 
the resonant $s$-channel amplitude for
$e^+e^- \to A'\to e^+e^-$ does not interfere with the analogous QED
process with an off-shell $\gamma$, nor with $t$-channel amplitudes
that can then be neglected.  Using the narrow width approximation
$\sigma_{\rm res}$  can be written as:
\begin{equation}
\label{eq:resxsection}
 \sigma_{\rm res}(E_e)  = \sigma_{\rm peak}
 \frac{\Gamma^2_{A'}/4}{(\sqrt{s}-m_{A'})^2+
\Gamma^2_{A'}/4}  \,, 
\end{equation} 
with $s\simeq 2 m_eE_e$, $ \sigma_{\rm peak} \simeq 12\pi/m^2_{A'}$
and $\Gamma_{A'} \simeq \epsilon^2 \alpha m_{A'}/3$.  In the numerical
computation we take into account $m_e$ effects both in the cross
section and in the width, and we also account for the emission of soft
photons from the initial state (see e.g.~\cite{Bohm:1982hr}) up to
energies $\Delta E/E_b \approx 1\%$, which can radiatively enhance the
resonance width, and thus the production rate.
%
\begin{figure}[t!!]
\begin{center}
\includegraphics[width=1\linewidth]{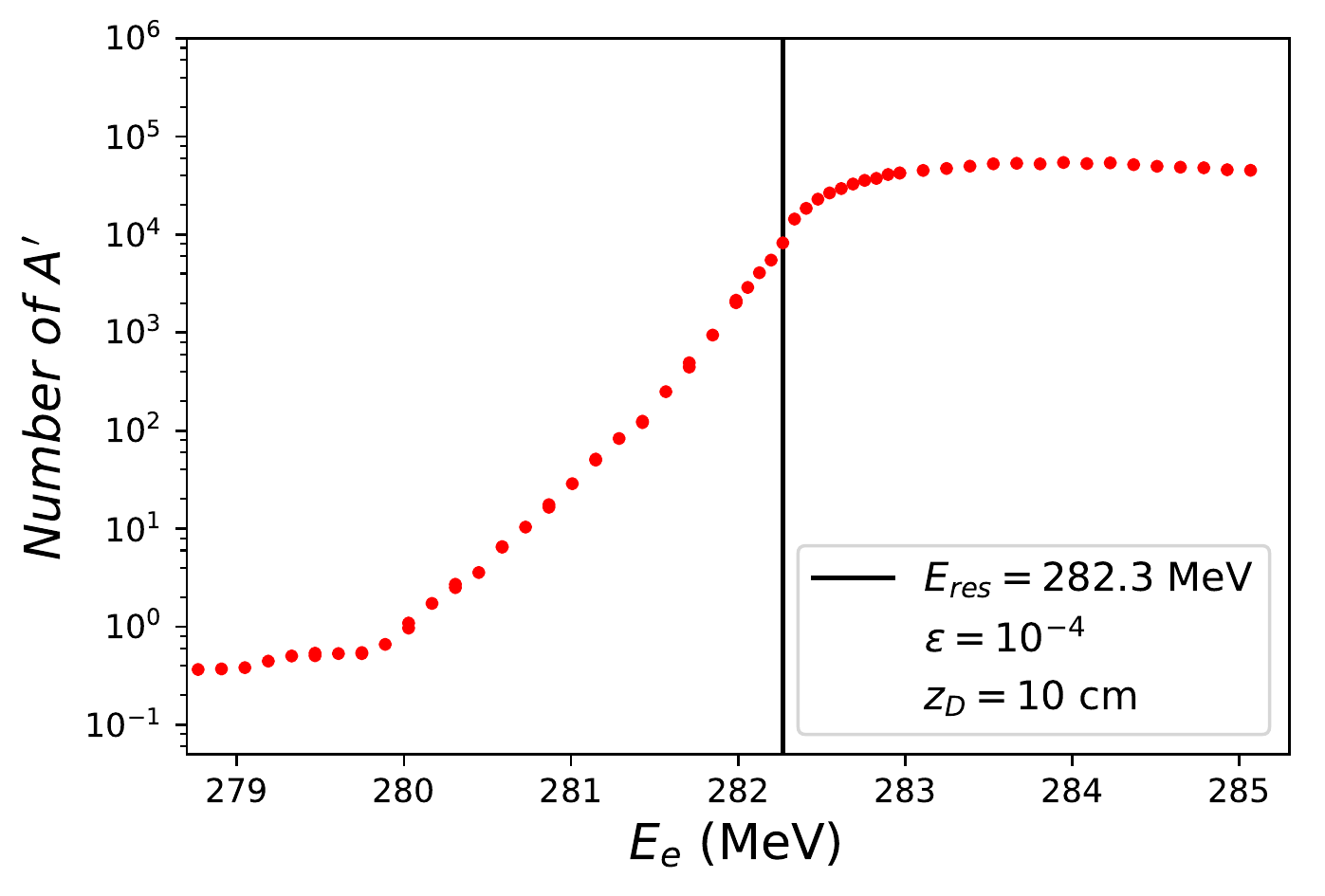}
\caption{The number of DP decaying outside the dump as a function of
  the beam energy for $\epsilon=10^{-4}$.  The vertical line
  corresponds to the energy for resonant production of a 17\,MeV DP.
  A dump length $z_D=10\,$cm and a background free measurement have
  been assumed.}
\label{fig:NDPvsE}
 \end{center}
\end{figure}
%
%
With respect to other DP production mechanisms, resonant production 
has some peculiarities and advantages: 

(i) The peak cross section does not depend on $\epsilon$ and the
dependence of the total resonant cross section is only quadratic
($\sim \epsilon^2\alpha$). As regards the observable number of electron-positron pairs
from $A'$ decays, for small $\epsilon$ the suppression in production
is over-compensated by the strong enhancement from the larger decay length
$\sim \exp(-\epsilon^2)$ which increases  the number of $A'$ that decay outside
the dump.  For this reason, resonant DP production in thick target
experiments is particularly well suited to explore the parameter space
at small $\epsilon$.

(ii) At fixed value of $\epsilon$, the $A'$ decay length
$\ell_\epsilon = \gamma\, c\, \tau_{A'}$ is independent of the value
of the $A'$ mass.  This is because $m_{A'}$ cancels between the boost
factor $\gamma \sim m_{A'}/(2 m_e)$ and the lifetime
$\tau_{A'} \propto 1/m_{A'}$. For all $A'$ masses the decay length is
then fixed $\ell_\epsilon\sim 3/(2m_e\alpha\epsilon^2)$. Therefore,
the entire $m_{A'}$ range within the reach of the beam energy can be
probed with the same sensitivity.

(iii) Under the reasonable assumption that the background remains
constant when the beam energy is varied by only a few MeV, the
background can be directly measured from the data.  This is
illustrated in \Fig{NDPvsE}: when the beam energy lies well below the
resonance, the background for $e^+e^-$ pairs (assumed to be absent for
the case of $z_D=10\,$cm in the picture) can be directly measured.
When the beam energy is increased, in approaching resonant production
the number of $e^+e^-$ pairs {\it produced} increases in a step-wise
way up to a maximum, and then remains approximately constant with
increasing energy, due to positron energy losses in the material,
which drive their energy towards $E_{\rm res}$.  Clearly, even in the
presence of a significant number $N_{BG}$ of $e^+e^-$ background
pairs, as long as $N_{A'} > \sqrt{N_{BG}}$ a signal of $A'$ decays can
be detected.\footnote{Such a spectacular signature would be
    prevented if the $A'$ resonance lies somewhat below the minimum
    beam energy, since one would always measure $e^+e^-$ resonantly
    produced by primary $e^+$ degraded in energy, together with
    backgrounds (we thank the referee for this remark). However, in
    this case by raising the beam energy and stepping further away
    from the resonance, the number of dilepton pairs resonantly
    produced would drop because of the degradation of the primary beam
    quality due to EM showering. The behavior of a `background' which
    decreases with increasing beam energy would still be a signal of
    beyond the SM physics.}

\begin{figure}[t!]
\begin{center}
\includegraphics[width=1\linewidth]{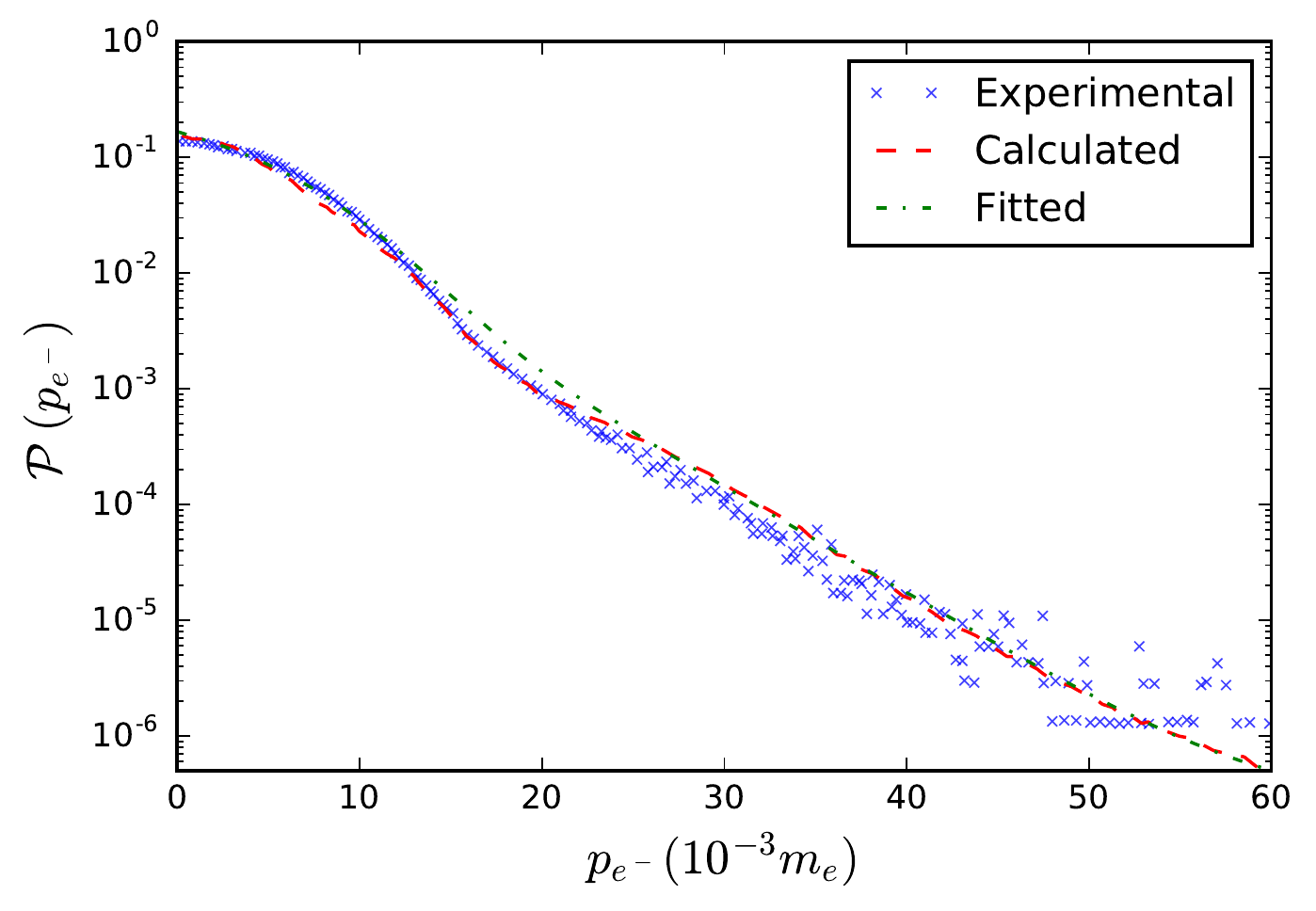}
\caption{The positron annihilation probability as a function of the
  target electron momentum for tungsten (figure adapted from
  Ref.~\cite{Ghosh:2000}).  The blue crosses represent experimental
  points, while the red dashed line is the result of the calculation
  method adopted in~\cite{Ghosh:2000}. The green dot-dashed line
  corresponds to the fit given by the function in
  \Eq{eq:Pdistribution}.  }
\label{fig:fit}
 \end{center}
\end{figure}
%


\begin{table}[t]
\setlength{\tabcolsep}{5pt} \global\long\def\arraystretch{1.3}
 \centering
 \begin{tabular}{|c|c|c|c|c|}
 \hline
 $\epsilon \Big{/} N_{A'}^{\rm prod}$ &$E_{\rm res}$ ($v_{e}=0$) &  $E_{\rm res}$ & $E_{\rm res}+2\sigma_b$ \\
 \hline
 \hline
 $1.0\times10^{-3}$ & $7.69\times10^{11}$ & $1.51\times10^{11} $ & $4.72\times10^{11} $  \\
 \hline
 $5.0\times10^{-4}$ & $1.81\times10^{11}$ & $3.79\times10^{10} $ & $1.17\times10^{11} $  \\
 \hline
 $1.0\times10^{-4}$ & $7.25\times10^{9}$ & $1.49\times10^{9} $ & $4.73\times10^{9} $  \\
 \hline
 \end{tabular}
 \caption{Number of $17\,$MeV DP produced in the first radiation length 
   of a tungsten target for $10^{18}$ positrons on target, for three different values of $\epsilon$.
   The second and third columns are for a  beam energy tuned to the resonant  
   value $E_{\rm res}= 282.3\,$MeV,  assuming respectively    
   electron at rest and with the velocity distribution  
   in~\Eq{eq:Pdistribution}. The last column, also  
   including $v_e$ effects, is for a  beam energy  $E_b = E_{\rm res}+2\sigma_b$. 
 }
\label{tab:NoDP}
\end{table}

\section{Effects of  target electrons velocities}

Inside materials electrons are not at rest, and in the case of large
atomic numbers, like tungsten $_{74}$W, electrons can have large
velocities, especially the ones in the inner core shells.  This can be easily
verified by estimating the electrons virial velocities
$\langle v_{nl}\rangle\approx \alpha Z^{(nl)}_{\rm eff}$ in terms of
the effective nuclear charge $Z^{(nl)}_{\rm eff}$ felt by electrons in
the $(nl)$ shell (a complete list of effective nuclear charges can be
found in Ref.~\cite{ClementiRaimondiReinhardt}).  For targets of small
atomic number, like $_{6}$C or $_{13}$Al, virial velocities are small,
and the effects of target electrons motion is likely to be negligible.
However, for
$_{74}$W one finds that the average velocities span a rather large
range $0.003 \lsim \langle v_{nl}\rangle \lsim 0.5$ when going from
valence or conduction electrons (with Fermi energy
$\epsilon_F\sim 4.5\,$eV) to inner core electrons.  Thus, for positron
annihilation in tungsten the center of mass (c.m.) energy can differ
sizeably from what can be naively estimated in terms of the beam
energy, energy spread, energy loss due to in-matter propagation, and
assuming electrons at rest. To give an example, already for a
longitudinal velocity component $v_z \sim 0.03$ the effect of shifting
the c.m. energy away from the resonant
value 
is three time larger than the effect of the intrinsic $\sim 1\%$
energy spread in the beam energy.  Of course, what is needed to
account for the c.m. energy shift is not simply the momentum
distribution of electrons, but rather the positron annihilation
probability as a function of the electron momentum, since
annihilation with de-localized and weakly bound valence electrons,
which contribute to the low-momentum part of the momentum
distribution, is more likely than annihilation with the localized and
tightly bound core electrons contributing to the high-momentum part.

For positron annihilation at rest, the annihilation probability
distribution as a function of the electron momentum is directly
measured from the Doppler broadening by the amount $\Delta E= p_L /2$
of the 511 keV photon line, with $p_L$ the $e^-$ momentum component
along the direction of $\gamma$ emission (the relative direction of
the two $\gamma$'s also deviates from $180^o$ by the small angle
$\theta =p_L /m_e$).  In \Fig{fit} (adapted from \cite{Ghosh:2000}) a
large set of experimental points for $_{74}$W is represented with blue
crosses. The red dashed line represents a theoretical calculation
performed in the same paper.  Up to
$p_{e^-} \sim 15\cdot 10^{-3}\,m_e$ the main contribution to the
annihilation comes from electrons in the 5d shell, beyond that point 4f
electrons dominate, while the contribution of the high momentum core
electrons becomes relevant only for
$p_{e^-} \gsim 40\cdot 10^{-3}\,m_e$ where, however, the annihilation
probability is suppressed below $10^{-5}$.  Accordingly, we find that a
good fit to the experimental and calculated
distributions~\cite{Ghosh:2000} can be obtained with the sum of just
three terms:
  \begin{equation} 
\label{eq:Pdistribution}
  \mathcal{P}(v_e)=\frac{1}{N}\Big(1.015^{-v_e^2}+1.112^{-2v_e}+
\theta(v_e-40)\,3\cdot10^{-6+\frac{1}{v_e}}\Big),
  \end{equation}
  where $v_e =p_{e^-}/m_e$, $N\sim 12$ is a normalization factor, and
  the first term in parenthesis accounts for 5d electrons, the second
  for 4f electrons, and the last one, which is non zero only for
  $v_e\geq 40$, accounts for core electrons.  To take into account
  target electron motion we thus replace the Mandelstam variable $s$
  in $\sigma_{\rm res}$ by
\begin{equation}
\label{eq:sve}
s(v_e,\chi) = 
2 m_e \left[E_e  \left(1-
\mathcal{P}(v_e) v_e \, \frac{1}{2}s_\chi c_\chi
\right)+m_e\right]\,,  
\end{equation}
where $c_\chi =\cos\chi$ accounts for the projection of $\vec v_e$
along the $z$-direction of the incoming positron, $s_\chi/2$ with
$s_\chi=\sin\chi$ is the probability distribution for the angle
$\chi$, and we integrate the cross section in $c_\chi$ and
$v_e \in [0,0.06]$.  \Tab{NoDP} collects some results that illustrate
how the number of DP {\em produced} within the first radiation length
of tungsten depends on various effects.  The second column gives the
results for three different values of $\epsilon$ for a beam energy
tuned at the resonant energy $E_{\rm res}=282.3\,$MeV, when the motion
of the target electrons is neglected.  The third column gives the
results obtained when the electron velocity is taken into account
according to the distribution in~\Eq{eq:Pdistribution}. We see that
the shift of the c.m. energy due to the electron momentum has the
effect of reducing the number of DP produced by about a factor of
five.  The last column gives the results for a beam energy tuned above
the resonance $E_b=E_{\rm res}+2\sigma_b$. The number of DP is
increased by about a factor of three because of the positron energy
losses, which brings on resonance also positrons in the high energy
tail of the initial energy distribution.

Of course, using the annihilation probability distribution for
positrons at rest in the problem at hand, is a crude way of
proceeding. We can expect that target electron motion effects can be
more sizeable for in-flight annihilation of short wavelength positrons
with energies of $O(100\,{\rm MeV})$, since the annihilation
probability with electrons in the inner shells will be enhanced.
Therefore, our estimate of the production rates might be optimistic by
a factor of a few. On the other hand, while positron energy loss, which
proceed mainly via bremsstrahlung, constitute a quantized process, the
dependence of the c.m. energy on the angle $\chi$ characterizing the
electron momentum is continuous, and this justifies modeling positron
energy losses as a continuous process.

%
\begin{figure*}[t!]
\begin{center}
\includegraphics[width=0.90\linewidth]{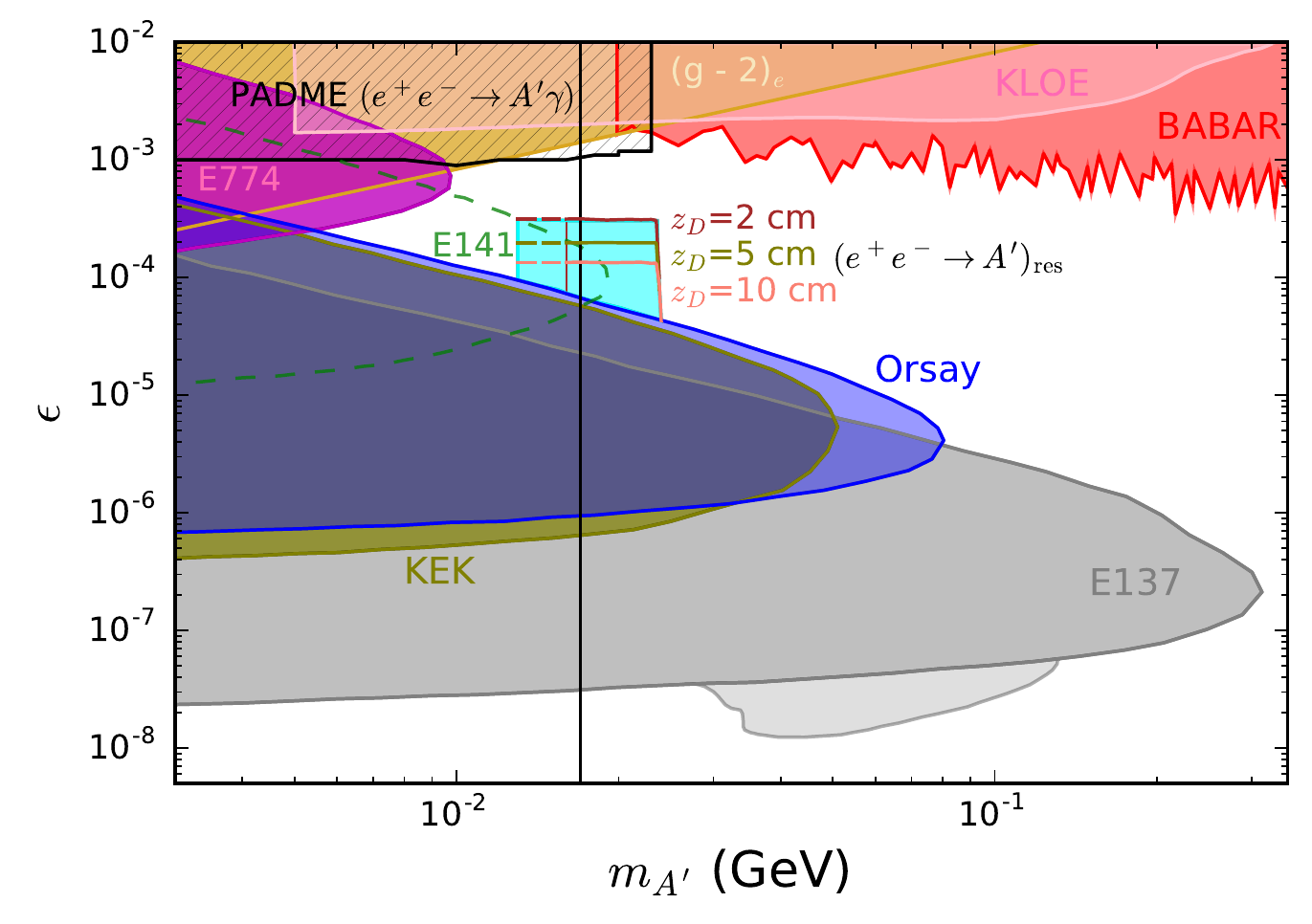}
\caption{Limits on the DP kinetic mixing $\epsilon$ as a function of the
  mass $m_{A'}$ from different experiments.  For 
  $m_{A'}\approx 17\,MeV$ (vertical black line) we consider still
  viable the region bounded from below by the Orsay and
  KEK blue and green-yellowish lines~\cite{Andreas:2012mt}
  and from above by the $(g-2)_e$ orange
  line~\cite{Endo:2012hp,Davoudiasl:2014kua}.    
  For reasons explained in the text we do not consider as firmly
  excluded the region around $m_{A'}\approx 17\,$MeV delimited by the
  black-dashed curve of the E141 SLAC
  experiment~\cite{Bjorken:2009mm,Andreas:2012mt}.  The region that
  could be excluded by PADME running in thin target mode is hatched in
  black, while the three trapezoidal-shaped areas give the PADME reach
  in thick target mode, respectively for a 10, 5 and 2\,cm tungsten
  dump, assuming zero background. These regions extend to $A'$ masses 
  lower than the mass corresponding to the minimum beam energy ($m_{A'}\sim
  16\,$MeV for $E_b^{\rm min} =250\,$MeV depicted with the thin brown
  vertical line) because of positron energy
  losses in propagating trough the material. 
The lower region in light gray
  extending the E137 exclusion limits is from the reanalysis in
  Ref.~\cite{Marsicano:2018krp}.}
\label{fig:limits}
 \end{center}
\end{figure*}
%

\section{Results}
\label{sec:results}

Before discussing the results a few words on backgrounds are in order.
The PADME spectrometers can detect $e^+e^-$ pairs with good resolution
for coincidence in time and momentum.  The $A'$ angular spread due to
the transverse momentum of atomic electrons is much less than the
intrinsic angular spread of the beam ($\sim 1 $ mrad) and it does not
affect the reconstruction of the coincidence.  For targets of
sufficient thickness, background from secondary $e^-$ detected in
coincidence with primary or secondary $e^+$ can be avoided by
measuring their depleted momentum via electromagnetic deflection.  For
targets of smaller length a certain number of $e^+e^-$ pairs retaining
a large fraction of the beam energy can exit the dump, and in this
case the data driven method of searching for a `knee' in the number of
$e^+e^-$ pairs versus beam energy (see \Fig{NDPvsE}) can provide a
precious tool for revealing the onset of resonant $e^+e^-$ production
on top of the background.  Punch-through photons, produced via
bremsstrahlung in the very first layers of the dump, carrying a large
fraction of the original beam energy, and converting in $e^+e^-$ in
the last millimeter or so, constitute the most dangerous background.
This background could be significantly suppressed by equipping the
experiment with a plastic scintillator veto few mm thick, or a silicon
detector of a few hundreds of $\mu$m, placed right at the end of the
dump, to ensure that the $e^+e^-$ pairs originate from decays in the
vacuum vessel outside the dump.  Additionally, if the experiment could
be equipped with a suitable tracker, able to provide an accurate
$e^+e^-$ invariant mass reconstruction, many sources of backgrounds
could be further reduced.  In particular, given that the invariant
mass of the $e^+e^-$ originating from photon conversion
$m^2_{e^+e^-}=0$ is very far from $m^2_{e^+e^-}\sim$(17\,MeV)$^2$
expected from resonant annihilation, the punch-through photon
background could be efficiently eliminated.

In \Fig{limits} we show the status of the current limits for DP
searches assuming visible $A'$ decays into $e^+e^-$ pairs with unit
branching fraction and suppressed couplings to the proton.  As is
discussed in Ref.~\cite{Feng:2016jff} the last assumption is required
in order to evade the tight constraints from $\pi^0\to \gamma A'$
obtained by the NA48/2 experiment~\cite{Batley:2015lha}, and to render
thus viable an explanation of the $^8$Be anomaly via an intermediate
$A'$ vector boson.  For this reason we have not included in
in~\Fig{limits} the limits from the NA48/2
experiment~\cite{Batley:2015lha} nor those from the $\nu$-Cal I
experiment at the U70 accelerator at IHEP
Serpukhov~\cite{Blumlein:2013cua,Blumlein:2011mv} which also do not
apply for protophobic $A'$.  In the figure, the vertical black line
gives the location of the DP resonance at $m_{A'}=17\,$MeV.  Leaving
aside the limits from the SLAC E141 experiment for which, as explained
in the introduction, the reach in $A'$ mass might be overestimated, a
viable window remains between the Orsay/KEK lines
($\epsilon \gsim 7\cdot 10^{-5}$) and the $(g-2)_e$ line
($\epsilon \lsim 1.4\cdot 10^{-3}$).  The black hatched region depicts
the forecasted sensitivity of PADME in thin target mode, that will
search for DP via the $e^+e^-\to A' \gamma$ process. The limits assume
$10^{13}\,$pot/yr.  The light cyan trapezoidal regions represent
instead the constraints that PADME could set by running in thick
target mode with $10^{18}\,$pot/yr, and are respectively for tungsten
targets of 10\,cm, 5\,cm and 2\,cm of length, and neglecting
backgrounds.
The BTF energy range for positron beams
$250\lsim E_b/{\rm MeV} \lsim 550$ corresponds to c.m. energies in the
interval $16\lsim E_{\rm c.m.}/{\rm MeV}\lsim 23.7$.  Neglecting a
possible small c.m. energy increase from target electron velocities,
the upper value sets the upper limit on the $A'$ masses that can be
produced. The lower c.m. energy limit is indicated by the thin
vertical brown line. However, because of positron energy losses, the
region at low $m_{A'}$ that can be explored extends to values smaller
than $16\,$MeV, as indicated in the figure. Of course, in propagating
well inside the dump, the beam gets degraded in energy, directions of
particle momenta, number of positrons, by several effects that we are
neglecting.  Therefore, we can expect that the experimental
sensitivity could be extended down to $m_{A'}$ values lower than
$16\,$MeV by no more than a few MeV. This might still be sufficient to
reach into the region where the E141 exclusion limits can be trusted.

In summary, it is apparent how the two PADME search modes are
complementary, since they can set new bounds respectively in the
regions of large $O(10^{-3})$ and small $O(10^{-4})$ values of the DP
mixing parameter $\epsilon$.  With some intense and dedicated
experimental efforts, the new regions in \Fig{limits} could be
explored in less than one year of running.  In particular, the allowed
window for the $^8$Be DP could be sizeably reduced, or its existence
could be unambiguously established.

\section{Conclusions}
In this letter we have suggested a new way to search for narrow
resonances, and specifically DP, coupled to $e^+ e^-$ pairs, via
resonant production in $e^+ e^-$ annihilation.  There are only a few
facilities around the world where positron beams in the 100 MeV - few
GeV range will be available. The Frascati BTF is one of those and it
can provide beams with energy between $250 - 550\,$MeV.
Coincidentally, this range covers precisely the c.m. energy needed to
produce via resonant $e^+e^-$ annihilation the $m_{A'} \sim 17\,$MeV
DP invoked to explain the anomaly observed in $^8$Be nuclear
transitions~\cite{Krasznahorkay:2015iga,Krasznahorkay:2017gwn,Krasznahorkay:2017qfd}.
By exploiting this production process, the Frascati PADME experiment, 
presently under commissioning, will be able to reach well
inside the interesting parameter space region.  \Fig{limits} shows
that a gap will remain between the large $\epsilon$ region that can be
bounded by searching for $A'$ produced via $e^+e^- \to \gamma A'$, and
the small $\epsilon$ region that can be efficiently explored via
resonant $e^+e^- \to A'$ production.  The reason for this gap is that
the first process, being of $O(\alpha^2 \epsilon^2)$, looses quickly
sensitivity when the value of $\epsilon$ is decreased too much, while
$A'$ production via resonant annihilation becomes inefficient when
$\epsilon$ becomes too large, so that most $A'\to e^+e^-$ decays occur
inside the dump.  Resonant $e^+e^- \to A'$ production is not relevant
for PADME running in thin target mode, because the large beam energy
$E_b \sim 550\,$MeV implies that positrons will always have energies
far from any narrow resonance with mass $ \lsim 23.7\,$MeV, given that
positron energy losses in the $100\,\mu$m diamond target are
negligible.  However, it is conceivable that by reducing the beam
energy down to $ \sim 282\,$MeV, by increasing the size of the
target to several $100\,\mu$m to enhance $A'$ resonant production, and
keeping the beam intensity well below $10^{18}$pot/yr to keep counting
rates inside the detector under control, at least part of the
remaining region for the $17\,$MeV DP could be explored, and maybe the
whole gap could be closed. We are presently exploring this
possibility.  Before concluding, we stress again that resonant
$e^+e^- \to A'$ production can be relevant also for electron 
beam dump experiments, since secondary positrons that could
trigger the annihilation process are abundantly produced in EM
showers. This feature has been recently exploited in reanalysing 
the SLAC E137 data~\cite{Marsicano:2018krp}, with the result of extending
the previously excluded region~\cite{Bjorken:2009mm,Andreas:2012mt}
towards smaller $\epsilon$ values, as is shown by the light gray area
in~\Fig{limits}.

\section*{Acknowledgments}
We thank M. Battaglieri, A. Celentano, V. Kozhuharov, L. Marsicano and
P. Valente for discussions. E.N. acknowledges enlightening conversations
with A. Arvanitaki, S. Dimopoulos, A. L. Morales Aramburo, M. Pospelov
and J. Pradler.  The work of E.N. was supported in part by the INFN
``Iniziativa Specifica'' Theoretical Astroparticle Physics (TAsP-LNF),
and by Perimeter Institute (PI) for Theoretical Physics. Research at PI
is supported by the Government of Canada through the Department of
Innovation, Science and Economic Development, and by the Province of
Ontario through the Ministry of Research, innovation and Science.
C.D.R.C. acknowledges support from COLCIENCIAS in Colombia (doctoral
scholarship 727-2015), and the LNF Theory Group for hospitality and
partial financial support during the development of this project.



\medskip


 \bibliographystyle{apsrev4-1.bst}
%


\end{document}